\RequirePackage{ifpdf}
\ifpdf
	\documentclass[onecolumn, prb, showkeys, letterpaper, pdftex]{revtex4}
\else
	\documentclass[onecolumn, prb, showkeys, letterpaper, dvips]{revtex4}
\fi

\ifpdf
  \usepackage[%
    pdftex,%
    pdftitle = {{Is diversity good?}},%
	pdfsubject = {{'Diversity' is ambiguous so that the lack of precise definition may lead to major errors. Diversity can have six meanings: it can be a factual description, a craving for symmetry, an intrinsic good, an instrumental good, a symptom, or a side effect.}},%
	pdfauthor = {{Mathieu Bouville}},%
	pdfkeywords = {{ethics, policy, higher education, affirmative action, minority students, female students}},%
	letterpaper,%
    hyperindex,%
    bookmarksopen,%
    bookmarksopenlevel=1
]{hyperref}
\else
  \usepackage[
	dvips,
    pdftitle = {{Is diversity good?}},%
	pdfsubject = {{'Diversity' is ambiguous so that the lack of precise definition may lead to major errors. Diversity can have six meanings: it can be a factual description, a craving for symmetry, an intrinsic good, an instrumental good, a symptom, or a side effect.}},%
	pdfauthor = {{Mathieu Bouville}},%
	pdfkeywords = {{ethics, policy, higher education, affirmative action, minority students, female students}},%
	letterpaper,%
    hyperindex,%
    bookmarksopen,%
    bookmarksopenlevel=1
]{hyperref}
\fi

\newcommand{\rb}[1]{\raisebox{-1.5ex}[0pt][0pt]{#1}}

\begin{document}

\addcontentsline{toc}{chapter}{Is diversity good?}
\title{Is diversity good?}

\author{Mathieu Bouville}
	\email{m-bouville@imre.a-star.edu.sg}
	\affiliation{Institute of Materials Research and Engineering, Singapore 117602}

\begin{abstract}
\addcontentsline{toc}{section}{Abstract}
Prominent ethical and policy issues such as affirmative action and female enrollment in science and engineering revolve around the idea that diversity is good. However, even though diversity is an ambiguous concept, a precise definition is seldom provided. We show that diversity may be construed as a factual description, a craving for symmetry, an intrinsic good, an instrumental good, a symptom, or a side effect. These acceptions differ vastly in their nature and properties. The first one cannot lead to any action and the second one is mistaken. Diversity as intrinsic good is a mere opinion, which cannot be concretely applied; moreover, the most commonly invoked forms of diversity (sexual and racial) are \emph{not} intrinsically good. On the other hand, diversity as instrumental good can be evaluated empirically and can give rise to policies, but these may be very weak. Finally, symptoms and side effects are not actually about diversity. We consider the example of female enrollment in science and engineering, interpreting the various arguments found in the literature in light of this polysemy.
\end{abstract}
\maketitle

\section{Introduction}
Diversity is at the core of widely discussed ethical and policy issues such as affirmative action in higher education,\cite{Crosby-06, Gurin-04, Orfield-01} female enrollment in science and engineering departments\cite{Chen-96, Baum-90, Wulf-diversity, Chubin-05, Lane-99, Widnall-88, Cohoon-02, Bouville-women-07, Bouville-phys_world-07, Anderson-02} ---``one of the greatest challenges facing this nation''\cite{Lane-99}---, diversity in the medical professions,\cite{Hammond-99, Cohen-97} workplace diversity,\cite{Griggs, handbook-diversity, Pless-04} and biodiversity.\cite{Tilman-00} However, in spite of its ``kudzu-like progression''\cite{Appiah-05} 
in recent years, diversity is seldom defined precisely.

As philosopher and Nobel laureate Bertrand Russell pointed out, ``everything is vague to a degree you do not realize till you have tried to make it precise.'' Who does not state precisely and explicitly why diversity should be increased cannot realize that `diversity' can have several meanings and is bound to confuse them. Believing that people who have the same name are the same person would seem ludicrous. It is nevertheless a very common error when it comes to diversity. 
For instance, to \citet{Pless-04} diversity is good for business because it ``can become a competitive advantage.'' They conclude that ``diversity is, first and foremost, a cultural question and thus a question of norms, values, beliefs and expectations. As such, it is an ethical question and determined by some very essential founding principles of human coexistence.'' As ``competitive advantage'' diversity is plainly an instrumental good (see Sec.~\ref{sec-instrumental}), whereas as ``ethical'' it is intrinsically good (Sec.~\ref{sec-intrinsic}) or perhaps a symptom (Sec.~\ref{sec-others}). These wholly different conceptions of diversity cannot be so equated.

The source of the problem is that issues related to diversity are seen as essentially political: people first decide whether they like diversity and then look for a rationalization. The meaning of diversity is drawn from the conclusion rather than the other way around. Paraphrasing Bradley, one could say that diversity is the finding of bad reasons for what we believe upon instinct. The merit of an argument then comes more from its capacity to support a pre-established conclusion than from its objective validity.

In contrast, we will first try and determine what diversity means and what its properties are before applying the concept (to the case of female enrollment in science and engineering, in Sec.~\ref{sec-women}).
This way we can talk of diversity clearly and unambiguously, and thus make sure that any policy based on diversity is justified and consistent.

\section{\label{sec-intrinsic}Diversity as intrinsic good}
It is sometimes claimed (and very often assumed) that diversity is intrinsically good. This means that diversity is good as an end, i.e.\ good in and for itself. One may for instance say that happiness, love, and friendship are intrinsically good.

\subsection{Is diversity intrinsically good?}
``Questions of ultimate ends are not amenable to direct proof.''\cite{Mill} Indeed, what kind of experiment or observation could prove diversity to be good in itself? Consequently, ``anything which is good as an end must be admitted to be good without proof.''\cite{principia_ethica} 
But of course there is no objective reason to leap into faith and admit without proof the opinion that diversity is an intrinsic good. 

In contrast, it is possible to show that diversity is not intrinsically good if the idea is inconsistent. What method can one use?
``That of considering what value we should attach to it if it existed in absolute isolation, stripped of all its usual accompaniments. And this is, in fact, the only method that can be safely used, when we wish to discover what degree of value a thing has in itself.''\cite{principia_ethica} 

If diversity were the only (or the most important) good then, if most people are smart and healthy, we would have to promote stupidity and sickness. We would also have to increase the number of women in prison, where they are under-represented.\cite{Bouville-women-07, Bouville-phys_world-07} Clearly, diversity cannot be the only or the supreme good: it may at most be one good amongst others. In fact, diversity may very well be good but of marginal importance (e.g.\ to be used as a tie-breaker if everything else is absolutely the same). But then the fact that something contributes to diversity would be a negligible detail, so that we could ignore diversity altogether. One cannot prove that diversity is important anymore than one can prove it to be intrinsically good.

Important note. `Not good' does not mean bad: something may be morally neutral or amoral. `Not intrinsically good' does not mean not good: something may be good without being intrinsically good.

\subsection{The scope of diversity}
Let us assume for the sake of argument that diversity is both intrinsically good and important. For practical purposes, we need to know exactly what one means by diversity, we need to know what kind of diversity we are talking about.
If diversity is defined very broadly, everybody contributes to it (everybody is different from the majority in at least one way), so that it is impossible to rely upon it to choose between people (for instance when hiring).

It seems that the most natural solution is to dismiss unimportant categories. But how to tell them apart? What makes skin color more important than hair color? Many would invoke a historical disadvantage or anti-discrimination laws to give more importance to the former. But if diversity is supposed to be \emph{intrinsically} good it cannot depend on history or the law. Another difference between skin and hair is that people pay much less attention to the latter. This makes diversity essentially about countering spontaneous behavior (see `diversity as symptom' below). But then the goodness of diversity is not intrinsic either (it cannot be both intrinsic and a reaction). The reasons why people focus on skin rather than hair are not compatible with the idea of an \emph{intrinsic} good.

Let us consider a man and a woman who have had essentially the same life. Verily, the woman will not contribute much more to diversity than the man since, apart from her sex, she is just like him: many men are more different from him than she is. 
One will reply that it is very unlikely that a man and a woman be so similar because they have a different ``life experience.''\cite{Wulf-diversity} This may be true, but what this means is that being a man or a woman is not directly relevant, it merely correlates with a form of diversity that is supposed to be good. Sexual difference ---``stripped of all its usual accompaniments''\cite{principia_ethica}--- is not good (or very marginally). Likewise racial diversity is not intrinsically good. If sexual and racial diversities are good, they are so only indirectly.

\subsection{From intrinsic good to policy}
Since it is logically impossible to prove that one kind of diversity or another is intrinsically good, the choice of scope is arbitrary, a personal belief. Yet, if one does not know what kind of diversity is intrinsically good (if any) one cannot know what kind of diversity to promote (if any). Any choice of policy would have to be the result of the bargaining between different viewpoints. Obviously what is supposed to be intrinsically good cannot logically be the result of a political compromise. Hence, no consensus can be reached for a policy based on diversity as intrinsic good (if there is a consensus then the consensual form of diversity is not intrinsically good): no such policy can concretely exist. 

One can also wonder how it is possible to agree that diversity is good and important without agreeing on what diversity is. If a Briton and an American both say `I like corn,' do they agree? No, because `corn' would not have the same meaning for the two of them. The agreement is a semantic illusion: they refer to different things for which they simply happen to use the same word. Likewise, without an agreement on the scope of diversity, any agreement that diversity is good is vacuous.

As already pointed out, the idea that diversity is intrinsically good is in itself arbitrary. Consequently, any policy is really based on an opinion, not on facts. It may be the opinion of the majority and this opinion may even become law, but neither can prove diversity to be a good in itself so that neither can justify a policy relying on diversity as intrinsic good.

\section{\label{sec-instrumental}Diversity as instrumental good}
\subsection{Introducing a new view of diversity as good}
What we saw in the previous section is not that diversity cannot be good but rather that \emph{intrinsic} goodness is too stringent a goal. 
In the present section, we introduce a different conception of diversity as good, which avoids the inconsistencies of intrinsic good. 

One says that diversity is good as a means (instrumental good) if it is conducive to something good. In the case of minorities in college, for instance, ``diversity has been shown to result in positive learning out\-comes and positive `democracy outcomes' for all students.''\cite{Crosby-06}
This was the only kind of argument accepted by the Supreme Court in \emph{Regents of the University of California v.\ Bakke}. More women in engineering would be good because of the greater variety of designs which more diverse teams could invent\cite{Wulf-diversity, Pless-04, Lane-99, Chubin-05} (``the innovative and creative potential inherent to a diverse workforce.'')\cite{Pless-04} 
To decide whether one should hire a new employee who is very smart but has the same skills as current employees or somebody who may be of average intelligence but with an original expertise, one compares the outcomes (which alternative leads to the best results?).

\subsection{Is diversity good and important?}
Saying that diversity is intrinsically good makes a statement about what \emph{is} good. Conversely, in the case of instrumental good, the good is independent of diversity: one can agree on what is good without making any claim related to diversity. There are then two separate questions: What is good? and Does diversity contribute to it? 
In engineering, a well designed product is good. To establish whether mixed groups lead to this good, one can compare the performances of all-male, all-female, and mixed groups. With sufficient data one could conclude beyond reasonable doubt whether diversity has good and important consequences. Likewise one can find out empirically whether diversity is important (leads to major improvement) or not (barely an improvement at all). That sexual diversity merely correlates with something genuinely good is not an issue here since we make no claim that sexual diversity is intrinsically good.

Anybody denying that diversity is good as means would then reject facts. With diversity as intrinsic good no proof can exist, so that denying that diversity is good is just rejecting an opinion.
Are opponents to diversity-based policies objectively wrong? Or do they simply have a different opinion? This cannot be answered until one has made clear in what diversity is supposed to be good.

\subsection{Diversity and ethics}
If diversity is good because it has good consequences, it is not universal: it is not diversity in general but some form of diversity that is good. Exactly what kind of diversity is good must be determined on a case-by-case basis.
It is important to notice that diversity is not necessarily a moral concept then: if more diversity leads to more money then diversity is a business concept for example. Diversity is ethical in nature only if it leads to a good which is itself ethical in nature.

An important consequence is that diversity then cannot be expected to receive a wide support. Indeed, if diversity is morally good then \emph{anybody} would be expected to contribute to increasing it, but if diversity is only a matter of making more money for a given company then people outside the company have no reason to try to increase diversity in this company. If one cannot show that diversity is a matter of ethics, its support will be very limited: only those who benefit directly have a reason to contribute. This is why proponents of affirmative action try to show that it benefits both blacks and whites:\cite{Crosby-06,Gurin-04} the whole population (and the courts) could not be expected to support it if it were good for minorities only.

\subsection{The problem of limited knowledge}
As we pointed out, any form of diversity is not necessarily conducive to the good, so that studies are needed to determine ---in every particular situation--- which ones are. If one then knows that a certain kind of people would be good for engineering, one may seek to increase their enrollment rather than that of other categories for which no benefit is \emph{known}. Yet, what one knows depends on what one wants to know. So that we favor a category because we decided to find out whether we should. What seemed neutral (favoring categories objectively conducive to the good) may in fact be dependent on human choices.

\section{\label{sec-others}Other acceptions of diversity}
We showed that diversity could be construed as an intrinsic good or as conducive to the good and that ``the nature of these two species of universal ethical judgments is extremely different; and a great part of the difficulties, which are met with in ordinary ethical speculation, are due to the failure to distinguish them clearly.''\cite{principia_ethica}
In this section we present four other ways to interpret diversity.

\subsection{Descriptive diversity}
Diversity may be simply descriptive. One can for example say that, if this state has 3\% of blacks and that state 30\%, the latter is more diverse. This greater diversity is a fact, not a judgment. One must notice that descriptive diversity is not sufficient for action. Three steps are necessary to decide to act. (i)~Learning about a fact in the external world. (ii)~Making a judgment on this neutral and objective fact that we know. (iii)~Deciding whether and how to act based on this judgment. Descriptive diversity is only step~i (the other five acceptions of diversity are meant to be judgments, step~ii): it is necessary but not sufficient.

\subsection{Diversity as symptom}
The symptoms of a disease neither are bad in themselves nor lead to anything bad: they are simply the \emph{consequence} of something bad.
Likewise, lack of diversity may not be wrong in itself, nor does it necessarily lead to anything bad. Rather, it can be seen as a hint that something is not right, as a symptom. Unlike when diversity is considered good, one does not purport to actively increase diversity: rather, the argument here is that diversity should not be artificially decreased (for instance by discrimination). Mentions of ``groups that have systematically faced discrimination and oppression''\cite{handbook-diversity} clearly interpret lack of diversity as a symptom.

One should notice that increasing diversity is not \emph{ipso facto} good in this case: getting rid of the symptom is not the same as treating the patient. For instance, arbitrarily expelling random white college students would increase diversity but would be utterly absurd. Moreover, lack of diversity is just a hint, not a proof. It is therefore necessary to look ---beyond the symptom--- at the good itself. In fact, as soon as one notices a problem, one should give up thinking in terms of diversity (i.e.\ of symptom) to focus on solving the actual problem. It is quite easy to lose focus and concentrate on a mere symptom.

\subsection{Diversity as a side effect}
According to \citet{Cohen-97}, ``seeking diversity in the medical professions is imperative to achieve just and equitable access to rewarding careers in the medical professions.'' Here the core of the argument is ``just and equitable access,'' not diversity. Diversity is merely a side effect in this case. Making this an argument in favor of diversity is misleading. Due to the new popularity of diversity (``it has become one of the most pious of the pieties of our age that the United States is a society of enormous cultural diversity''),\cite{Appiah-05} 
basing an argument on `diversity' gives it weight.
For the sake of clarity one should talk of increasing diversity only when the justifications are genuinely based on diversity, not just something correlated with it.

\subsection{A craving for symmetry}
Diversity is often construed as the idea that any subset of the population should be similar to the overall population: a group representing, say, 20\% of the overall population must have 20\% of representatives in board rooms, higher education, science, etc. It is quite difficult to understand in what the same proportion in a specific context as in the overall population is supposed to be good in itself. It rather seems a craving for symmetry. Yet, this is what most people have in mind what they talk of diversity.

\section{\label{sec-women}An example \texorpdfstring{---}{-} Female enrollment in science and engineering}
The six acceptions of diversity we identified (summed up in Table~\ref{table-summary}) can be found in the literature arguing for a greater female enrollment in science and engineering. As we will see, they are seldom clearly identified, so that what the authors intended to say is often unclear.

\begin{table*}
\begin{tabular}{lcccccc}
\hline
conception of~	& ~relation to~&\rb{~factual}
							& \rb{specific}&\rb{decisive~}&\rb{~ethical}&\rb{drawback(s)}\\
diversity		& the good 		& 		&		& 			&  			& \\
\hline
descriptive		& none 			& yes	& yes	& no		& no		& one cannot act upon it\\
symmetry		& mistaken 		& no	& no	&by mistake	& no		& mistaken\\
good in itself	& is 			& no	& no	& no		& yes		& mere opinion; inapplicable\\
good as means	& causes		& yes	& yes	& yes		& not in	& needs case-by-case proofs; \\
				& 		 		& 		& 		& 			& all cases	& may have limited support \\
symptom			& caused 		& yes	& yes	& indirectly& indirectly& indirect\\
side effect		& indirect 		& yes	& yes	& indirectly& indirectly& misleading\\
\hline
\end{tabular}
\caption{\label{table-summary} Summary of properties of different conceptions of diversity. Specific: is this view of diversity different for different cases or is it rather universal? Decisive: allowing to make concrete decisions.}
\end{table*}

\subsection{Diversity-based argument bric-a-brac}
``While women make up over 50\% of the college-age population in the U.S., they represent a small minority among engineering students.''\cite{Chen-96} If this is said from the viewpoint of descriptive diversity, we can only agree that this is a fact. Yet it is clear that allusions to under-representation cannot be meant to be purely descriptive since they are supposed to justify policies to increase female enrollment in scientific fields. And, unlike what \citet{Baum-90} believes (``the numbers speak for themselves, demonstrating a significant problem in recruiting and retaining women''), a description cannot on its own justify action. 
Clearly there must be an implicit intermediate step, which justifies diversity-based policies.

Authors may value ``diversity for its own sake,''\cite{Chubin-05} yet few actually explicitly say so. 
Or they may crave for symmetry. This seems to be the case of \citet{Lane-99} who talks of ``a workforce for the global economy that reflects our great diversity.'' 
\Citet{Chubin-05} write that engineering ``must narrow the gap between practitioners on the one hand, and their clientele on the other'' but also that ``diversity for its own sake may speak to morality and fairness'' and that diversity is ``an enabler that makes teams more creative.'' So they invoke diversity as intrinsic good, as instrumental good, and as symptom and they wish for symmetry. Four different conceptions of diversity, no less! 
\Citet{Anderson-02} want to ``increase diversity in engineering'' (so says their title) but never explain why diversity should be increased (they use the word `diversity' a grand total of twice).

\subsection{Instrumental diversity}
Some argue with \citet{Wulf-diversity} that more women in engineering would be good because of the greater variety of designs which more diverse teams could invent. This is a valid argument provided that (i)~one establishes empirically that more diverse engineering teams indeed perform better, (ii)~this improvement will benefit society as a whole rather than just some technological companies (otherwise one would have no reason to contribute to this increase in diversity, as already pointed out), and (iii)~any other group that can contribute as much as women is treated in the same way.

According to \citet{Gurin-04} ``students who interact with diverse students in classrooms and in the broad campus environment will be more motivated and better able to participate in a heterogeneous and complex society.'' This means that if a heterogeneous (i.e.\ diverse) society is good, then diversity in college is good. But this begs the question: it shows that diversity is good (in college) assuming that diversity is good (in general). Even thought it seems to hold diversity to be an instrumental good, this argument really assumes that it is intrinsically good.
(This does not make the argument of diversity as instrumental good itself invalid since there can be positive outcomes that do not assume that diversity is good in itself, e.g.\ ``enhanced cognition.'')\cite{Crosby-06}

\subsection{Lack of diversity as symptom}
It is often argued that the low female enrollment in engineering is a symptom of social pressure\cite{Cohoon-02, Widnall-88} and that ``we should of course seek to rid any biases, for instance due to a girl's family or from schoolteachers.''\cite{Bouville-phys_world-07} When ``a lot of people argue for diversity in terms of fairness,''\cite{Wulf-diversity} they likely interpret lack of diversity as a symptom of lack of fairness.

Yet many authors nevertheless talk of women as under-represented. This seems to indicate that they would expect women to make up 50\% of engineers, which is simply wishing for symmetry. 
However, if by `under-represented' one means that there are fewer women in science and engineering than there would be without social biases, one is indeed arguing that lack of diversity is a symptom of something bad.\cite{Bouville-women-07} But this kind of under-representation cannot be measured: as we said earlier, as soon as we think that, through a lack of diversity, we identified something bad, we should stop talking of the symptom and focus on the actual problem. Moreover one should make sure that lack of diversity as symptom \mbox{---which} can be a valid argument--- is not mistaken for an irrational longing for symmetry.

\subsection{Diversity as side effect}
A common argument is that technology plays an ever-increasing role in society and economy, so that the number of engineering graduates must increase. And women are a great potential source of engineers.\cite{Chen-96, Baum-90, Wulf-diversity, Chubin-05, Lane-99, Widnall-88} Obviously this argument is not in favor of diversity itself: it aims at increasing the total number of engineers rather than change the composition of the engineering student body. The latter may be altered in the process, but this is only a side effect.

\section{Conclusion}
Since many different concepts have been called `diversity,' one must be careful when invoking diversity not to ring a homonym. This is of the utmost importance since some of these views are mistaken and cannot justify any policy.  
We saw that diversity may be a neutral description; but this is insufficient to act. 
The idea that there should be the same representation in a specific context as in the overall population is both puzzling and arbitrary. 
One cannot prove that diversity is good in itself. In fact, we showed that sexual and racial diversities are \emph{not} intrinsically good. Moreover, there cannot concretely exist policies based on diversity as intrinsic good. 
On the other hand, diversity as instrumental good can be evaluated empirically. But this must be done in every separate context. Diversity as instrumental good can give rise to precise policies.  But, since it is not intrinsically ethical, there is no reason for these policies to be widely supported. 
While diversity as symptom and as side effect can justify actions, these arguments are not actually about diversity so that one should give up thinking in terms of diversity to focus on what really matters.
Naturally, these are not necessarily mutually exclusive, e.g.\ diversity may lead to something good and the current lack of diversity may be a symptom of something bad. But the various acceptions must be introduced as separate arguments and each must be justified.



\begin{thebibliography}{20}

\bibitem[{{Gurin et~al.}(2004)}]{Gurin-04}
Gurin, P., Nagda, B.~A., \& Lopez, G.~E. (2004). The benefits of diversity in education for democratic citizenship. \emph{Journal of Social Issues}, \textbf{60}, 17--34.
 
\bibitem[{Crosby et~al.(2006)}]{Crosby-06}
Crosby, F.~J., Iyer, A., \& Sincharoen, S. (2006). Understanding affirmative action. \emph{Annual Review of Psychology}, \textbf{57}, 585--611.


\bibitem[{{Orfield}(2001)}]{Orfield-01}
Orfield, G. (2001). \emph{Diversity Challenged: Evidence on the Impact of Affirmative Action}. Cambridge: Harvard Education Publishing Group.

\bibitem[{{Lane}(1999)}]{Lane-99}
Lane, N. (1999). Increasing diversity in the engineering workforce. \emph{The Bridge}, \textbf{29}(2), 15--19.

\bibitem[{{Wulf}(1998)}]{Wulf-diversity}
Wulf, W.~A. (1998). Diversity in engineering. \emph{The Bridge}, \textbf{28}(4), 8--13.

\bibitem[{{Chubin \emph{et~al.}}(2005)}]{Chubin-05}
Chubin, D.~E., May, G.~S., \& Babco, E.~L. (2005). Diversifying the engineering workforce. \emph{Journal of Engineering Education}, \textbf{94}, 73--86.

\bibitem[{{Chen \emph{et~al.}}(1996)}]{Chen-96}
Chen, J.~C., Owusu-Ofori, S., Pai, D., Toca-McDowell, E., Wang, \mbox{S.-L.}, \& Waters, C.~K. (1996). A study of female academic performance in Mechanical Engineering. \emph{Frontiers in Education Conference}. 
Available online at: \url{http://fie.engrng.pitt.edu/fie96/papers/276.pdf}.

\bibitem[{{Baum}(1990)}]{Baum-90}
Baum, E. (1990). Recruiting and graduating women --- The underrepresented student. \emph{IEEE Communications Magazine}, \textbf{28}, 47--50.

\bibitem[{{Widnall}(1988)}]{Widnall-88}
Widnall, S.~E. (1988). AAAS Presidential Lecture: Voices from the Pipeline. \emph{Science}, \textbf{241}, 1740--1745.

\bibitem[{{Bouville}(2007)}]{Bouville-phys_world-07}
Bouville, M. (2007). Should we enrol more female students in physics? \emph{Physics World}, \textbf{20}(4), 18.

\bibitem[{Bouville(submitted)}]{Bouville-women-07}
Bouville, M. Should there be more women in science and engineering? \emph{submitted} (e-print: \url{http://arxiv.org/physics/0611089}).

\bibitem[{{Anderson and Northwood}(2002)}]{Anderson-02}
Anderson, L. \& Northwood, D. (2002). Recruitment and retention programmes to increase diversity in engineering. \emph{International Conference on Engineering Education}. Available online at: \url{http://www.ineer.org/Events/ICEE2002/Proceedings/Papers/Index/O065-O070/O069.pdf}.

\bibitem[{{Cohoon}(2002)}]{Cohoon-02}
Cohoon, J.~M. (2002). Recruiting and retaining women in undergraduate computing majors. \emph{SIGCSE Bulletin}, \textbf{{34}}, {48--52}.

\bibitem[{{Hammond}(1999)}]{Hammond-99}
Hammond, R.~A. (1999). The moral imperatives for diversity. \emph{Clinical Orthopaedics and Related Research} \textbf{362}, 102--106.

\bibitem[{{Cohen}(1997)}]{Cohen-97}
Cohen, J.~J. (1997). Finishing the bridge to diversity. \emph{Academic Medicine}, \textbf{72}, 103--109.

\bibitem[{{Griggs and Louw}(1995)}]{Griggs}
Griggs, L.~B. \& Louw, L.-L. (1995). \emph{Valuing Diversity}. New York: McGraw-Hill.

\bibitem[{{\emph{Handbook of Workplace Diversity}}(2006)}]{handbook-diversity}
Konrad, A.~M., Prasad, P., \& Pringle, J.~K. (2006). \emph{Handbook of Workplace Diversity}. London: Sage.

\bibitem[{Pless and Maak(2004)}]{Pless-04}
Pless, N.~M. \& Maak, T. (2004). Building an inclusive diversity culture: principles, processes and practice. \emph{Journal of Business Ethics}, \textbf{54}, 129--147.

\bibitem[{{Tilman}(2000)}]{Tilman-00}
Tilman, D. (2000). Causes, consequences and ethics of biodiversity. \emph{Nature}, \textbf{405}, 208--211.


\bibitem[{{Appiah}(2005)}]{Appiah-05}
Appiah, K. A. (2005). \emph{The Ethics of Identity}. Princeton, NJ: Princeton University Press.

\bibitem[{{Mill}(1871)}]{Mill}
Mill, J.~S. (1871). \emph{Utilitarianism}.

\bibitem[{G.~E. Moore(1903)}]{principia_ethica}
Moore, G.~E. (1903). \emph{Principia Ethica}. Cambridge: Cambridge University Press.

\end{thebibliography}
\end{document}